\documentclass[letterpaper]{JHEP3}
\usepackage{epsfig}

\newcommand{\roughly}[1]{\mathrel{\raise.3ex\hbox{$#1$\kern-0.85em
\lower1ex\hbox{$\sim$}}}}

\newcommand{\lsim}{\roughly<}
\newcommand{\gsim}{\roughly>}

\def\exd{{\hbox{d}}}

\def\ba{\begin{eqnarray}}
\def\ea{\end{eqnarray}}
\def\be{\begin{equation}}
\def\ee{\end{equation}}

\def\ssH{{\scriptscriptstyle H}}

\def\EF{{\scriptscriptstyle EF}}

\def\A{\mathcal{A}}

\def\E{\mathcal{E}}

\def\H{\mathcal{H}}

\def\L{\mathcal{L}}
\def\O{\mathcal{O}}

\def\eff{{\rm eff}}

\def\nn{\nonumber}

\def\({\left(}
\def\){\right)}

\def\pref#1{(\ref{#1})}

\title{Power-counting and the Validity of the Classical Approximation
 During Inflation}

\author{C.P. Burgess,${}^{1-2}$ Hyun Min Lee,${}^2$ and
Michael Trott${}^1$ \\
${}^1$ Perimeter Institute for Theoretical Physics, Waterloo ON,
N2L 2Y5, Canada.\\
${}^2$ Dept. of Physics \& Astronomy, McMaster University,
Hamilton ON, L8S 4M1, Canada.
}

\date{}

\abstract {We use the power-counting formalism of effective field
theory to study the size of loop corrections in theories of
slow-roll inflation, with the aim of more precisely identifying the
limits of validity of the usual classical inflationary treatments.
We keep our analysis as general as possible in order to
systematically identify the most important corrections to the
classical inflaton dynamics. Although most slow-roll models lie
within the semiclassical domain, we find the consistency of the
Higgs-Inflaton scenario to be more delicate due to the proximity
between the Hubble scale during inflation and the upper bound
allowed by unitarity on the new-physics scale associated with the
breakdown of the semiclassical approximation within the effective
theory. Similar remarks apply to curvature-squared inflationary
models.}

\begin{document}

\section{Introduction}
The hypothesis that the universe underwent accelerated expansion
during an early inflationary epoch can explain the flatness,
isotropy, homogeneity, horizon and undesired relic problems of the
early universe. Typically, inflation is caused by a local Lorentz
invariant energy density  dominating the equation of state and driving
an exponential expansion of the comoving Hubble length
\cite{firstINF}. Even better, the growth of quantum fluctuations
during inflation allows a simple description of the observed
features of the primordial cosmological fluctuations that are
required in the Hot Big Bang to seed the large-scale structure
observed in the universe. The general predictions of inflationary
scenarios also agree with the increasingly precise observations of
the properties of the Cosmic Microwave Background (CMB), such as
measured most recently by WMAP \cite{deltaval}. While the idea of
inflation is in good qualitative and quantitative agreement with
the data, it has so far proven more difficult to embed inflation
within a more complete framework of physics at the very high
energies that are required.

Thus, many inflationary scenarios exist that are constructed to be
consistent with the current experimental constraints. The vast
majority of these fall into the category of `slow-roll' inflation,
for which a scalar field (inflaton), classically evolves under the
influence of a very flat potential. It is the approximately
constant energy density of the scalar during this classical slow
roll that drives the inflationary epoch. Although some scenarios
are more sophisticated, and have incorporated important quantum
effects modifying or generating the potential, the usage of the
semi-classical approximation is standard in inflation studies.

In this paper, we apply the power counting formalism of effective
field theory to study the question of the size of the loop
corrections of the scalars coupled to gravity that are commonly
employed in the inflation literature. Our results are constructed
to be as general as possible, are not limited to one loop, and
allow one to directly examine the quantum corrections of physical
quantities (like the classical inflaton potential or scattering
cross sections). The techniques used rely on simple dimensional
arguments that are known to work for similar applications of
non-renormalizable theories in non-gravitational situations (like
chiral perturbation theory in the strong interactions). The beauty
of the approach is its simplicity, since the constraints on
couplings and masses that underlie the validity of the
semi-classical approximation can be quickly determined using
power-counting arguments without the need for extensive explicit
calculation (and yet agrees with these calculations when they are
available).

As an example of the utility of the formalism we develop, we study
the unusually predictive and simple Higgs-Inflaton scenario
\cite{Hinfa}. In this scenario, it is the Standard Model's Higgs
boson itself that acts as the inflaton, a scenario that is made
possible through the addition of the single dimension-four
interaction, $\delta \L = \xi H^\dagger H \, R$, that is usually
neglected, but that is expected to be {\em required} to exist due
to renormalization of the theory in curved space \cite{CShiggs}.
This term encodes the experimentally untested possibility of a
large nonminimal coupling of the Higgs to gravity, and the freedom
to choose the new coupling $\xi$ is all that is required to ensure
an inflationary slow roll.

However, these conclusions are drawn using a semi-classical
analysis, and we show that the domain of validity of this
approximation is very narrow for this model due to the large size of
$\xi \simeq 10^4$ required for successful inflation (consistent with WMAP constraints). We find
that the semiclassical analysis requires that the scale $M$, defining
the limit of validity of the effective theory, lies in the narrow
window $M_p/\xi \gg M \gg \sqrt{\lambda_\ssH} M_p/\xi$, where
$\lambda_\ssH$ is the usual quartic self-coupling of the Higgs in
the Standard Model potential. We show how this condition is very
sensitive to the existence of other heavy particles in the
microscopic theory that couple to the Higgs, even if these
couplings are quite weak. Similar remarks apply to
curvature-squared inflationary models, which also walk a thin line
of consistency.

\section{Power-counting}

Power counting the scales that appear in loops is a standard
technique of effective field theory, for which many excellent
reviews exist \cite{EFTreview} in the literature, including
applications to gravity \cite{GREFT,GREFT1}. In this section, we
use power counting to identify how successive terms in the
semiclassical expansion depend on the various scales and couplings
of the inflationary theory of interest. There are two types of
effective field theories normally considered in the literature for
inflation, that differ according to whether or not they focus on
the complete inflaton-metric system \cite{GREFTInfa}, or on the
specific adiabatic mode which (for single-field models) controls
the spectrum of primordial perturbations \cite{GREFTInfb}. We here
consider theories of the form of \cite{GREFTInfa} (and its
multi-scalar generalizations), and provide a power-counting
analysis of the order in the low-energy expansion at which any
effective interaction contributes.

\subsection{The effective field theory}

For definiteness, consider the following effective lagrangian,
describing the low-energy interactions of $N$ dimensionless scalar
fields, $\theta^i$, and the metric, $g_{\mu\nu}$:
\ba \label{Leffdef}
 - \frac{ \L_\eff}{\sqrt{-g}} &=& v^4 V(\theta) + \frac{M_p^2}{2}
 \, g^{\mu\nu} \Bigl[  W(\theta) \, R_{\mu\nu}
 + G_{ij}(\theta) \, \partial_\mu \theta^i
 \partial_\nu \theta^j \Bigr] \\
 && \quad + A(\theta) (\partial \theta)^4 + B(\theta)
 \, R^2 + C(\theta) \, R \, (\partial \theta)^2
 + \frac{E(\theta)}{M^2} \, (\partial \theta)^6
 + \frac{F(\theta)}{M^2} \, R^3 + \cdots \,. \nn
\ea
Here the lagrangian is organized as a derivative expansion, with
terms involving up to two derivatives written explicitly and the
rest only written schematically in order to sketch the dimension
of the coefficients. In particular $R^3$ collectively represents
all possible independent invariants constructed from three Riemann
tensors, or two Riemann tensors and two of its covariant
derivatives; $R (\partial \theta)^2$ denotes all possible
invariants involving one power of the Riemann tensor and two
derivatives acting on $\theta^i$; and so on for the other terms.

In eq.~\pref{Leffdef} the scalar fields are normalized so that the
coefficient of their kinetic terms is the reduced Planck mass,
defined in terms of Newton's constant by\footnote{This
normalization is convenient for large-field inflationary models,
for which the scalars move over Planckian distances in field
space, but we also consider scalars whose couplings are stronger
than Planck-suppressed in what follows below by including
couplings that carry compensating powers of $M_p$.} $M_p = (8\pi
G)^{-1/2}$. All of the coefficient functions, $V(\theta)$,
$G_{ij}(\theta)$, $A(\theta)$ and so on, are dimensionless, and
the scale $M$ that makes up the dimensions is taken to be
characteristic of whatever underlying microscopic physics has been
integrated out.\footnote{That is, $M$ might be regarded as the
lightest of the particles that were integrated out to produce the
low-energy theory. Our calculations below show why such a mass
would appear in this way.} Since it is the {\em smallest} mass
that dominates in such a denominator, it is important to recognize
that generically $M \ll M_p$ \cite{GREFT}. In applications to
inflation our interest is usually (but not always) in situations
where $V \simeq v^4 \ll M^4$ when $\theta \simeq \O(1)$.

For the purposes of estimating the size of quantum effects, we
expand about a classical solution,
\be
 \theta^i(x) = \vartheta^i(x) + \frac{\phi^i(x)}{M_p}
 \quad \hbox{and} \quad
 g_{\mu\nu} (x) = \hat g_{\mu\nu} (x) +
 \frac{h_{\mu\nu}(x)}{M_p} \,,
\ee
which allows the effective action, eq.~\pref{Leffdef}, to be
written as a sum of effective interactions
\be \label{Leffphih}
 \L_\eff = \hat \L_\eff + M^2 M_p^2 \sum_{n}
 \frac{c_{n}}{M^{d_{n}}} \; \O_{n} \left(
 \frac{\phi}{M_p} , \frac{ h_{\mu\nu}}{M_p} \right)
\ee
where $\hat\L_\eff = \L_\eff(\vartheta,\hat g_{\mu\nu})$ is the
lagrangian density evaluated at the background configuration.
The sum over $n$ runs over the labels for a complete set of
interactions, $\O_{n}$, each of which involves $N_n = N^{(\phi)}_n
+ N^{(h)}_n \ge 2$ powers of the fields $\phi^i$ and $h_{\mu\nu}$.
($N_n \ne 1$ follows as a consequence of the background field
equations for $\vartheta^i$ and $\hat g_{\mu\nu}$.) The parameter
$d_{n}$ counts the number of derivatives appearing in $\O_n$, and
so the factor $M^{-d_n}$ is what is required to keep the
coefficients, $c_n$, dimensionless. The overall prefactor, $M^2
M_p^2$, is chosen so that the kinetic terms --- {\it i.e.} those
terms in the sum for which $d_n = N_n = 2$ --- are $M$ and
$M_p$ independent. Notice also that the operators $\O_n$ depend
implicitly on the properties of the classical backgrounds,
$\vartheta^i$ and $\hat g_{\mu\nu}$, about which the expansion is
performed.

Comparing eqs.~\pref{Leffdef} and \pref{Leffphih} also shows that
there are factors of the scales $v$, $M$ and $M_p$ buried in the
dimensionless coefficients $c_n$. In particular, any term
involving no derivatives comes from the scalar potential,
$V(\theta)$, and so
\be \label{cndeq0}
 c_n = \left( \frac{v^4}{M^2 M_p^2} \right) \lambda_n
 \qquad \hbox{(if $d_n = 0$)} \,,
\ee
where the $\lambda_n$ represent dimensionless couplings that are
independent of $M_p$ and $M$. Similarly, the absence of $M_p$ in
all of the terms involving more than two derivatives in
eq.~\pref{Leffdef} implies
\be \label{cndgt2}
 c_n = \left( \frac{ M^2}{ M_p^2} \right) g_n
 \qquad \hbox{(if $d_n > 2$)} \,,
\ee
where $g_n$ is similarly independent of $M$ and $M_p$.

In terms of the $\lambda_n$'s the scalar potential has the
schematic form
\be
 V(\phi) = v^4 \left[ \lambda_0 + \lambda_2 \left(
 \frac{\phi}{M_p} \right)^2 + \lambda_4 \left(
 \frac{\phi}{M_p} \right)^4 + \cdots \right] \,,
\ee
which shows that the natural scale for the scalar masses under the
above assumptions is $m \simeq v^2/M_p$. The quartic coupling
constant, $\lambda_4 (v/M_p)^4$, is similarly Planck suppressed.
Such small masses and couplings follow from the assumption that
$V$ only runs through a range of order $v^4$ as $\phi$ runs all
the way out to $M_p$. Although such a shallow potential often
arises in inflationary applications, in some circumstances it is
also interesting to consider potentials for which $V \sim v^4$
when $\phi$ runs over a comparable range, $\phi \sim v$, and so
for which $m \simeq v$ up to dimensionless couplings. Such
potentials can be included in the above analysis by further
redefining
\be \label{checklambda}
 \lambda_n = \left( \frac{M_p}{v} \right)^{\check N_n} \, \check
 \lambda_n \,,
\ee
in the power-counting rules that are to follow. Here $\check N_n
\le N_n$ denotes the number of scalar fields of this type
appearing in the vertex in question. This need not agree with
$N_n$ if there are also other scalars, or graviton vertices,
appearing in the $d_n = 0$ vertex of interest.

\subsection{Semiclassical perturbation theory}

Our goal is to follow how the couplings $c_n$ and the scales $M$
and $M_p$ appear in physical quantities at various orders of the
semiclassical expansion. To this end we divide $\L_\eff$ into an
unperturbed and perturbed lagrangian density,
\be
 \L_\eff = \Bigl( \hat \L_\eff + \L_0 \Bigr)
 + \L_{\rm int} \,,
\ee
where $\L_0$ consists of those terms in $\L_\eff$ for which $N_n =
2$ and $d_n \le 2$. Since the path integral over $\phi^i$ and
$h_{\mu\nu}$ is Gaussian in the absence of $\L_{\rm int}$ we can
define the semiclassical expansion in principle by computing the
generator, $\Gamma$, of 1-particle irreducible (1PI) graphs
perturbatively in $\L_{\rm int}$. This is a semiclassical
expansion because the leading contribution is the classical result
\be
 \Gamma[\theta, g_{\mu\nu}] = \int \exd^4 x \;
 \L_\eff(\theta, g_{\mu\nu}) + \cdots\,.
\ee

The key issue is to identify what the small quantity is that makes
such an expansion a good approximation. To determine this, imagine
now computing a contribution to $\Gamma$ coming from a Feynman
graph involving $\E$ external lines. The propagators, $G(x,y)$,
associated with each of the $I$ internal lines in this graph come
from inverting the differential operator that is defined by the
term $\L_0$. The important thing about these for the present
purposes is that they do not depend on $M$ and $M_p$, although
they can depend on scales (like the Hubble scale, $H$) that appear
in the background configurations, $\vartheta^i$ and $\hat
g_{\mu\nu}$.

Similarly, vertices in this graph all come from terms in $\L_{\rm
int}$, and so each time the interaction $\O_n$ contributes a
vertex to the graph it comes with a factor of $c_n M_p^{2 - N_n}
M^{2-d_n}$. If the graph contains a total of $V_n$ such vertices
it acquires in this way a factor
\be
 \prod_n \Bigl[ c_n M_p^{2 - N_n} M^{2 - d_n} \Bigr]^{V_n}
 = M_p^{2 - 2L - \E} \, \prod_n \Bigl[  c_n M^{2 - d_n}
 \Bigr]^{V_n}  \,,
\ee
where the equality uses the identity
\be \label{endcons}
 2I + \E = \sum_n N_n V_n
\ee
that expresses that the end of each line in the graph must occur
at a vertex, as well as the definition,
\be \label{loopdef}
 L =1 + I - \sum_n V_n  \,,
\ee
of the number of loops, $L$, of the graph.

\subsubsection*{Power-counting}

The relative contribution of each graph to $\Gamma$ is then
simplest to enumerate using dimensional arguments. However any
such argument is complicated by the ultraviolet divergences
that arise in the integration over the positions, $x$, of the
vertices; divergences that may be traced to the singularities in
the propagators, $G(x,y)$, in the coincidence limit $y \to x$. For
the purposes of making the dimensional argument it is therefore
simplest to regularize these divergences using dimensional
regularization, since in this case all of the dimensions of the
various integrations is set by a physical scale appearing in the
problem (such as the masses of the fields $\phi^i$, or a scale
like $H$ characterizing the size of a derivative of the background
classical configuration).\footnote{Naively, using a cutoff to
regulate these divergences would seem to change the estimates we
are about to make. However the cutoff-dependent estimates found in
this way are guaranteed to cancel cutoff-dependent counter-terms
once the theory is renormalized, since physical quantities cannot
depend on how we choose to arbitrarily regulate a graph. What
counts physically is how observables depend on observable (or
renormalized) quantities, and using a cutoff regularization simply
makes it difficult to follow dimensional analysis through
intermediate steps of the calculation. Of course the final answer
does not depend on how the calculation is performed, and any
strong dependence on a cutoff in the regularized theory shows up
in dimensional regularization as a dependence on a large physical
scale in the problem, such as the mass of a heavy particle that
has been integrated out.}

Suppose now that $E$ denotes the largest of the physical scales
that appear explicitly in the propagators or vertices of the
calculation. Then to leading approximation we can neglect any
other, smaller, scales compared with $E$ when estimating the size
of a particular Feynman graph. Since the contributions to $\Gamma$
all share the same dimension as the initial lagrangian density
$\L_\eff$, the contribution of a graph involving $\E$ external
lines, $L$ loops and $V_n$ vertices involving $d_n$ derivatives
becomes\footnote{For simple backgrounds these calculations can be
made in momentum space (although the dimensional argument being
made does not require this), and when this is done the reader
should note that eq.~\pref{PCresult0} pulls out the standard
overall momentum-conserving factor, $(2 \, \pi)^4 \, \delta^4(q)$,
from $ \A_\E (E)$, where $q$ denotes the total 4-momentum flowing
into the graph.}
\be \label{PCresult0}
 \A_\E (E) \simeq E^2 M_p^2 \left( \frac{1}{M_p} \right)^\E
 \left( \frac{E}{4 \pi \, M_p}
 \right)^{2L} \prod_n \left[ c_n \left( \frac{E}{M}
 \right)^{d_n-2} \right]^{V_n} \,.
\ee
The factors of $4\pi$ in this expression come from standard
arguments. (For example, for a flat background with constant
$\vartheta^i$, they arise from the loop-integral measure in
momentum space, $\int {\exd^4 p}/{(2\pi)^4}$, once the angular
integration over the momentum direction is taken into account.)

Keeping in mind the factors of $M$ and $M_p$ that are hidden in
some of the $c_n$'s --- {\it c.f.} eqs.~\pref{cndeq0} and
\pref{cndgt2} --- it is useful to write separately the terms with
$d_n = 0$ and $d_n = 2$ in the product, to get
\ba \label{PCresult}
 \A_\E (E) &\simeq& E^2 M_p^2 \left( \frac{1}{M_p} \right)^\E
 \left( \frac{E}{4 \pi \, M_p}
 \right)^{2L} \prod_{d_n = 2} \Bigl( c_n \Bigr)^{V_n} \\
 && \qquad \qquad \qquad \times
 \prod_{d_n = 0} \left[ \lambda_n \left( \frac{v^4}{E^2 M_p^2}
 \right) \right]^{V_n}
 \prod_{d_n \ge 4} \left[ g_n \left( \frac{E}{M_p}
 \right)^2 \left( \frac{E}{M}
 \right)^{d_n-4} \right]^{V_n}  \,. \nn
\ea

Eq.~\pref{PCresult} is the main result of this section. It shows
in particular what combination of scales must be small in order to
justify the validity of the perturbative expansion. A generic
sufficient condition for successive insertions of interactions to
be smaller than preceding ones is to have $E$ be sufficiently
small,
\be
 \frac{E}{4 \pi M_p} \ll 1 \,,
\ee
and
\be
 g_n \left( \frac{E}{M_p}
 \right)^2 \left( \frac{E}{M}
 \right)^{d_n-4} \ll 1 \quad \hbox{(for $d_n \ge 4$)}\,.
\ee

Repeated insertions of two-derivative interactions do not
generically generate large contributions provided
\be
 c_n \ll 1 \quad \hbox{(for $d_n = 2$)} \,,
\ee
although having $c_n \simeq \O(1)$ need not cause problems if
symmetries strongly constrain the kinds of interactions of this
kind that can arise. For example, for pure gravity only the
Einstein-Hilbert action itself has two derivatives, for which all
the resulting graviton interactions have $c_n$'s of order one. The
lack of suppression of these interactions shows that they are all
generically equally important in a given low-energy
process.\footnote{That is, although the low-energy expansion
controls higher derivatives, for generic relativistic applications
in General Relativity one must work to all orders in the expansion
of the metric about a given background, $g_{\mu\nu} = \overline
g_{\mu\nu} + h_{\mu\nu}$.}

Finally, the only place where inverse powers of $E$ arise is
associated with no-derivative interactions, and {\it a-priori}
these seem like they could be dangerous in a low-energy expansion
since
\be
 \lambda_n \left( \frac{v^4}{E^2 M_p^2}
 \right) \ll 1 \quad \hbox{(for $d_n = 0$)}
\ee
might not be satisfied. This would be even more worrisome in the
event that the potential has the form $V = v^4 f(\phi/v)$ for some
order-one function $f(x)$, since in this case we have seen ---
{\it c.f.} eq.~\pref{checklambda} --- that we must take $\lambda_n
= (M_p/v)^{\check N_n} \check \lambda_n$ for vertices involving
these scalars. Although it is true that low-energy is not itself
sufficient to suppress these interactions, their presence need not
destroy the low-energy approximation due to correlations that the
topology of a graph imposes amongst the numbers of loops, the
number of vertices and the number of external lines, as we now
see.

For example, imagine the potential worst-case scenario for the
low-energy expansion where all of the vertices of the Feynman
graph have $d_n = 0$ and $\check N_n = N_n$ ({\it i.e.} only
involve the largest and most dangerous couplings). In this case
the identities \pref{endcons} and \pref{loopdef} hold separately
for the internal lines and vertices involving only the dangerous
scalar, and so
\be \label{checkID}
 \sum_n (\check N_n - 2) \check V_n = \check \E - 2 + 2L \,,
\ee
leading to
\ba \label{PCresultcheck}
 \A_{\check \E} (E) &\simeq& E^2 M_p^2
 \left( \frac{1}{M_p} \right)^{\check \E}
 \left( \frac{E}{4 \pi \, M_p} \right)^{2L}
 \prod_{d_n = 0} \left[ \check \lambda_n \left(
 \frac{M_p}{v} \right)^{\check N_n}
 \left( \frac{v^4}{E^2 M_p^2} \right) \right]^{\check V_n} \nn\\
 &\simeq& E^2 v^2
 \left( \frac{1}{v} \right)^{\check \E}
 \left( \frac{E}{4 \pi \, v} \right)^{2L}
 \prod_{d_n = 0} \left[ \check \lambda_n
 \left( \frac{v^2}{E^2} \right) \right]^{\check V_n} \,.
\ea
Clearly all powers of $M_p$ have dropped out in this expression
and, as we see in more detail below, provided $\phi \simeq E$ the
net power of $E/v$ that appears in $\Gamma$ is then
\be
 \left( \frac{E}{v} \right)^{2 + \check \E + 2L - \sum_n 2\check
 V_n} = \left( \frac{E}{v} \right)^{4 + \sum_n(\check N_n - 4)
 \check V_n} \,,
\ee
which uses eq.~\pref{checkID} once more. This shows that quintic
and higher interactions generate only positive powers of $E/v$,
while quartic interactions are neither enhanced nor suppressed by
$E/v$ (and so must be controlled purely by the small size of the
relevant dimensionless couplings, $\check \lambda_n$).

Since there are no interactions with $N_n = 1$ (by virtue of the
background field equations) or $N_n = 2$ (as these are `mass'
terms in the unperturbed lagrangian density), only the
super-renormalizable cubic terms with $N_n = 3$ are potentially
dangerous to the low-energy expansion (unless their dimensionless
coefficients are also suppressed so that $\check \lambda_3 \simeq
\O(E/v)$). Such trilinear vertices can indeed cause trouble for
the low-energy expansion, if they are of order $\lambda_3 v^4
(\phi/M_p)^3 \simeq \check \lambda_3 v \, \phi^3$, since $v$ need
not be small compared with the low-energy scales, $E$, to which
the effective theory is applied.

\subsection{Examples}
\label{ss:examples}

Eq.~\pref{PCresult} has a number of interesting special cases.

\subsubsection*{Pure gravity with no cosmological constant}

The only thing in the above arguments to change in the case of
pure gravity ({\it i.e.} no scalar fields) in the absence of a
cosmological constant is the absence of interactions having $d_n =
0$. In this case eq.~\pref{PCresult} reproduces the standard
result for General Relativity \cite{GREFT}. It predicts, in
particular, that for any $\E$ the dominant contributions arise for
$L=0$ with only vertices satisfying $d_n = 2$ included. For pure
gravity these graphs amount to working with General Relativity in
the purely classical limit. The first sub-leading contributions
may be similarly found, and correspond to working with General
Relativity at one loop ({\it i.e.} with $L=1$ and $V_n = 0$ unless
$d_n = 2$), or working at classical level and allowing precisely
one insertion from a curvature-squared interaction ({\it i.e.}
with $L=0$ and $V_n = 0$ for $d_n > 4$, $V_n = 1$ for $d_n = 4$
and $V_n$ arbitrary if $d_n = 2$).

\subsubsection*{Integrating out a particle of mass $m \ll M$}

Another application specializes to the case where the largest
scale in the amplitude is the mass, $m$, of a particle that is
being integrated out. In this case provided all other scales are
much smaller than $m$ the result for the $\Gamma$ is local, and
expression \pref{PCresult0} or \pref{PCresult} can be regarded as
describing how effective interactions are renormalized in
$\L_\eff$ due to the removal of this particle. More
quantitatively, if
\be
 V(\theta) = v^4 \left[ \lambda_0 + \lambda_2 \theta^2 +
 \lambda_4 \theta^4 + \cdots \right]
 = v^4 \, \lambda_0 + \frac{\lambda_2 v^4}{M_p^2} \, \phi^2 +
 \frac{\lambda_4 v^4}{M_p^4} \, \phi^4 + \cdots \,,
\ee
with all $\lambda_n$'s being of order unity, then the masses of
the $\theta^i$ particles are of order $m \simeq v^2/M_p$. To make
one particle systematically heavy relative to the others, we
either require $\lambda_2 \gg 1$ for the heavy field (as above,
where $\lambda_2 = (M_p/v)^2 \check \lambda_2$, say) or $\lambda_2
\ll 1$ for all of the others.

Since the largest scale in the Feynman graphs is $m$ by
assumption, we may use the above power-counting estimates with $E
\simeq m$. Furthermore, if we focus on contributions to $\A_\E$
that involve precisely $D$ derivatives, denoted $\A_\E^D$, then
the same dimensional arguments as above predict the following
scaling:
\ba \label{Renresult0}
 \A_\E^D &\simeq& m^2 M_p^2 \left( \frac{\partial}{m} \right)^D
 \left( \frac{1}{M_p} \right)^\E
 \left( \frac{m}{4 \pi \, M_p}
 \right)^{2L} \prod_n \left[ c_n \left( \frac{m}{M}
 \right)^{d_n-2} \right]^{V_n} \nn\\
 &\simeq& m^2 M_p^2 \left( \frac{\partial}{m} \right)^D
 \left( \frac{1}{M_p} \right)^\E
 \left( \frac{m}{4 \pi \, M_p}
 \right)^{2L} \prod_{d_n = 2} \Bigl( c_n \Bigr)^{V_n} \\
 && \qquad \qquad \qquad \times
 \prod_{d_n = 0} \left[ \lambda_n \left( \frac{v^4}{m^2 M_p^2}
 \right) \right]^{V_n}
 \prod_{d_n \ge 4} \left[ g_n \left( \frac{m}{M_p}
 \right)^2 \left( \frac{m}{M}
 \right)^{d_n-4} \right]^{V_n}  \,.\nn
\ea

Comparing this with the coefficients of the effective interaction
valid below the scale $m$, defined using the form of
eq.~\pref{Leffphih} (but with $M$ replaced by $m$)
\be \label{Leffphihtilde}
 \tilde\L_\eff = \hat \L_\eff + m^2 M_p^2 \sum_{n}
 \frac{\tilde c_{n}}{m^{\tilde d_{n}}} \; \O_{n} \left(
 \frac{\phi}{M_p} , \frac{ h_{\mu\nu}}{M_p} \right) \,,
\ee
we see the Feynman graph in question contributes
\be
 \delta \tilde c_n \simeq \left( \frac{m}{4 \pi \, M_p}
 \right)^{2L} \prod_n \left[ c_n \left( \frac{m}{M}
 \right)^{d_n-2} \right]^{V_n} \,.
\ee

In terms of the $m$- and $M_p$-independent, dimensionless
couplings, $\tilde \lambda_n$, $\tilde g_n$, $\lambda_n$ and
$g_n$, these become
\be
 \delta \tilde \lambda_n \simeq \left(
 \frac{m^2 M_p^2}{v^4} \right)
 \left( \frac{m}{4 \pi \, M_p}
 \right)^{2L} \prod_{d_n = 2} \Bigl( c_n \Bigr)^{V_n}
 \prod_{d_n = 0} \left[ \lambda_n \left( \frac{v^4}{m^2 M_p^2}
 \right) \right]^{V_n}
 \prod_{d_n \ge 4} \left[ g_n \left( \frac{m}{M_p}
 \right)^2 \left( \frac{m}{M}
 \right)^{d_n-4} \right]^{V_n}
\ee
while for $\tilde d_n \ge 4$ we instead have
\be
 \delta \tilde g_n \simeq \left( \frac{M_p^2}{m^2} \right)
 \left( \frac{m}{4 \pi \, M_p}
 \right)^{2L} \prod_{d_n = 2} \Bigl( c_n \Bigr)^{V_n}
 \prod_{d_n = 0} \left[ \lambda_n \left( \frac{v^4}{m^2 M_p^2}
 \right) \right]^{V_n}
 \prod_{d_n \ge 4} \left[ g_n \left( \frac{m}{M_p}
 \right)^2 \left( \frac{m}{M}
 \right)^{d_n-4} \right]^{V_n}  \,.
\ee

For example, at tree level ($L=0$) the corrections to couplings in
the scalar potential ($\tilde d_n = 0$) are of order
\be \label{deltalambdatree}
 \delta \tilde \lambda_n \simeq
 \left( \frac{m^2 M_p^2}{v^4} \right)
 \prod_n \left[ \lambda_n \left( \frac{v^4}{m^2 M_p^2}
 \right) \right]^{V_n}  \,,
\ee
because at tree level only $d_n = 0$ vertices can contribute to an
effective interaction having $\tilde d_n = 0$. Expanding this
tree-level result in graphs involving one, two and more vertices
then gives\footnote{Notice that one-particle reducible graphs are
allowed to contribute to the low-energy effective action, which is
only required to be irreducible with respect to the cutting of
{\em light} particle lines.}
\be \label{extree}
 \tilde \lambda_n \simeq \lambda_n +
 \left( \frac{v^4}{m^2 M_p^2} \right)  \sum_{\rm graphs}
  k_{mn} \, \lambda_n \lambda_m + \cdots \,,
\ee
where $k_{mn}$ are calculable coefficients and the sum is over
graphs for which $d_n = d_m = 0$ and $N_n + N_m = \tilde N_n + 2$.
The ellipses indicate tree level graphs involving three or more
vertices. Similarly, one-loop graphs involving only one vertex
contribute (for $\tilde d_n = 0$),
\be \label{exloop}
 \delta \tilde \lambda_n \simeq \frac{1}{(4 \pi)^2}
 \sum_n \left\{ r_n c_n \left( \frac{m^4}{v^4} \right) +
 \frac{m^2}{M_p^2} \left[ s_n \lambda_n
 + t_n g_n \left( \frac{m^4}{v^4}
 \right) \left( \frac{m}{M}
 \right)^{d_n-4} \right] \right\} \,,
\ee
where $r_n$, $s_n$ and $t_n$ are calculable, and so on.

Notice that if $m \simeq v^2/M_p$ then $m/M_p \simeq (v/M_p)^2 \ll
m/v \simeq v/M_p \ll 1$, and $m^2 M_p^2 \simeq v^4$. This
implies no suppression by scales between the terms in
eq.~\pref{extree}, while in eq.~\pref{exloop} it makes the sums
involving $c_n$ and $\lambda_n$ of the same order as one another,
but larger than those involving $g_n$. On the other hand, if $m$
is dialled up to $m \simeq v$, such as by taking $\lambda_n \simeq
(M_p/v)^{N_n}$ for some vertices in the scalar potential, then the
factor in eq.~\pref{extree} becomes $(v^4/m^2 M_p^2)(M_p/v)^{N_m +
N_n - \tilde N_n} = (v^2/M_p^2)(M_p/v)^{N_m + N_n - \tilde N_n} =
\O(1)$. In the loop expression, however, it is the $\lambda_n$
term that dominates (unsuppressed by powers of $v/M_p$) for
corrections to the $(M_p/v)$-enhanced couplings, while the $c_n$
and $\lambda_n$ terms compete (again unsuppressed by $v/M_p$) for
the corrections to the generic $\lambda_n$'s. In all cases the
$g_n$ coupling is subdominant.

\section{Applications to Inflation}

In applications to slow-roll inflation the background fields are
time-dependent, and so among the important scales in the problem
are the characteristic times over which the various fields vary
appreciably. For the metric this is given by the Hubble scale
\be
 H = \frac{\dot a}{a} \simeq \frac{\sqrt{V}}{M_p}
 \simeq \frac{v^2}{M_p} \,,
\ee
while the evolution of the inflationary scalar is similarly
characterized by the scale
\be
 \mu_\phi = \frac{\dot \phi}{\phi} \,.
\ee

During slow-roll inflation the scales $\mu_\phi$ and $H$ are
related to one another by the slow-roll conditions, which state
that the inflaton time derivative satisfies
\be \label{SReqn}
 \dot\phi \simeq \frac{V'}{H} \simeq \frac{M_p V'}{\sqrt{V}}
 \simeq \sqrt{\epsilon V} \simeq \sqrt\epsilon \; v^2\,.
\ee
Here
\be
 \epsilon = \frac12 \left( \frac{M_p V'}{V} \right)^2 \quad
 \hbox{and} \quad
 \eta = \frac{M_p^2 V''}{V} \,,
\ee
are the two slow-roll parameters \cite{slowroll}, where the
derivatives are taken with respect to the canonically normalized
fields. They arise because a necessary condition
for a slow roll is that both must be small: $\epsilon, |\eta| \ll
1$. Eq.~\pref{SReqn} implies that during a slow roll the relative
size of $H$ and $\mu_\phi$, depends on the size of $\phi$, with
\be
 \mu_\phi =
 \frac{\dot\phi}{\phi} \simeq \frac{\sqrt\epsilon \, v^2}{M_p} \simeq
 \sqrt\epsilon \, H \quad \hbox{if $\phi \simeq M_p$}\,,\qquad
 \mu_\phi =
 \frac{\dot \phi}{\phi} \simeq \sqrt\epsilon \, v \quad
 \hbox{if $\phi \simeq v$} \,.
\ee

The observation that the inflaton-gravity action is a part of the
more general effective lagrangian, eq.~\pref{Leffdef}, imposes
often unspoken conditions on the domain of validity of any
analysis that bases inflation on its classical solutions. It
requires in particular that the inflationary motion must be
adiabatic, which puts an upper limit on the inflationary
time-scales: $\mu_\phi, H \ll M$. Indeed, regarding the effective
theory as a derivative expansion breaks down if $H, \mu_\phi
\simeq M$, because then terms involving powers of $R/M^2 \simeq
(H/M)^2$ or $(\partial \theta)^2/M^2 \simeq (\mu_\phi/M)^2$ are
not small.

For many inflationary models there is an important constraint that
restricts the freedom to choose $H$ and $\mu_\phi$ arbitrarily.
This constraint arises when primordial fluctuations are regarded
as arising as quantum fluctuations of the inflaton during
inflation. Agreement with the observed temperature fluctuations in
the CMB requires the amplitude of curvature perturbations to have
a specific amplitude $\Delta_{\mathcal{R}}^2 |_{k^\star} = 2.445
\, \pm 0.096 \times 10^{-9}$ \cite{deltaval}, where $k^\star =
0.002 \, {\rm Mpc^{-1}}$.

When these perturbations are generated by quantum fluctuations in
$\phi$, then the quantity that controls their amplitude is $\delta  = H^2/\dot \phi
 = (24 \, \pi^2 \Delta_{\mathcal{R}}^2 |_{k^\star})^{1/2}$,
and so using the above estimates for $\dot \phi$ and $H$ gives
\be
 \delta \simeq \frac{1}{\sqrt{\epsilon}} \; \left(
 \frac{v}{M_p} \right)^2 \simeq 7 \times 10^{-4} \,.
\ee
This provides the important relationship $v/M_p \simeq 0.03 \,
\epsilon^{1/4}$.

\subsection{Corrections to inflationary scenaria}

Eq.~\pref{PCresult} allows an estimate of how the various
effective interactions contribute to an inflationary scenario,
provided $E$ is chosen to be the largest scale in the problem.

\subsubsection*{Classical effects from higher effective interactions}

The first modification to consider is the contribution of the
various effective interactions in eq.~\pref{Leffdef} to the
classical equations of motion. In the language of the estimate
\pref{PCresult} this amounts to asking the relative size of
various contributions in the classical limit ({\it i.e.} when
$L=0$). Eq.~\pref{PCresult} shows that (provided $g_n \lsim
\O(1)$) higher-derivative interactions with $d_n \ge 4$ are
suppressed by at least two powers of $E/M_p$, plus additional
powers of $E/M$ if $d_n > 4$. On the other hand, interactions with
$d_n = 2$ are not particularly suppressed, and generically neither
are interactions from the scalar potential. These two quantities
must therefore be included exactly into the classical calculation.
In particular, it is often a bad approximation to work in the
small-field limit that is implicit when expanding the potential in
powers of $\phi$, and neglecting terms beyond a particular power
(like quartic) when in the inflationary regime, as has recently
been re-emphasized within the context of string theory
\cite{phi6}.

\subsubsection*{Quantum contributions}

A second question asks about the size of quantum corrections to
the classical approximation. The size of these effects depends
crucially on how massive are the particles whose quantum
fluctuations are under study. In all cases eq.~\pref{PCresult}
applies (or \pref{PCresultcheck} if the natural scale for $\phi$
is $\phi \simeq v$ rather than $\phi \simeq M_p$), with $E \simeq
m$ for quantum fluctuations from particles whose mass satisfies $m
\gg \mu_\phi$, $H$, while $E \simeq \hbox{max}(\mu_\phi, H)$ for
the quantum effects of particles satisfying $m \ll \mu_\phi$, $H$.

%\pagebreak
\medskip\noindent{\it Heavy particles:}

\medskip\noindent
The limit $E \simeq m \gg H \simeq v^2/M_p$ leads to the estimates
of section \ref{ss:examples}, with the additional information that
$v^4/(E^2M_p^2) \simeq v^4/(m^2 M_p^2) \simeq H^2/m^2 \ll 1$. This
shows that in addition to the generic loop factor $(m / 4 \pi
M_p)^2$, the $d_n \ge 4$ interactions --- $g_n$ --- are further
suppressed by at least two powers of $m^2/M_p^2$, and interactions
in the scalar potential --- $\lambda_n$ --- are additionally
suppressed by powers of $H^2/m^2$. Only the $d_n = 2$ interactions
--- $c_n$ --- remain unsuppressed beyond the basic loop factor if
$\lambda_n \lsim {\cal O}(1)$. On the other hand, if there are
interactions in the scalar potential that are unsuppressed by
powers of $M_p$ (such as if $\lambda_n \simeq (M_p/v)^{N_n} \check
\lambda_n$, as discussed above) then loops involving these
interactions can also modify the inflaton mass in a dangerous way
\cite{EFTadiabatic2}.

Provided the heavy field itself only moves adiabatically, the
implications of loop effects of this type are most simply seen by
integrating the particle out, leading again to an effective theory
of the form of eq.~\pref{Leffdef}, but with $M$ replaced by $m$
\cite{EFTadiabatic1,EFTadiabatic2,TP,TASIReview}. As we have seen,
only those interactions having two or fewer derivatives
generically have an appreciable influence on the classical
equations, since the effects of interactions with $d_n \ge 4$ have
been argued already to be small. In general, quantum corrections
can change the shape of the classical potential, and such changes
can ruin the inflationary slow roll of the original potential
unless they are absorbed into the coefficients of the coefficients
of the original effective action. This is particularly true when
$\phi$ arises in the scalar potential $V$ suppressed by a light
scale like $v$ rather than $M_p$. This simply represents the usual
naturalness problems in keeping low-dimension terms in the scalar
potential small as heavier particles are integrated
out.\footnote{Approximate symmetries, such as shift symmetries
\cite{NatInf}, can protect the size of such corrections, although
it is important that these symmetries apply to {\it all} couplings
of the inflaton and not just to the self-couplings that appear in
the inflaton potential.}

Unfortunately, although these corrections need not be small, and
can undermine whether or not we believe a given theory actually
exhibits inflation in the first place, they do not have {\it
observable} implications in the sense that cosmological
observations are unable to separate quantum from classical
contributions to the potential. On the other hand, if the
heavy-field motion is not adiabatic, it need not decouple and so
cannot be integrated out. In this case its presence can generate
observable deviations from standard inflationary predictions
\cite{EFTnonadiabatic}.

\medskip\noindent{\it Light particles:}

\medskip\noindent
The analysis is different when the mass of the particle in the
loop is small compared with $H$ and $\mu_\phi$, which includes in
particular the inflaton itself since its mass is $m^2 = V'' = \eta
V/M_p^2 \simeq \eta \, v^4/M_p^2 \simeq \eta H^2 \ll H^2$. In this
case the estimate \pref{PCresult} still applies, but it is $E
\simeq H$ (since $\mu_\phi \simeq \sqrt\epsilon \, H \ll H$ in
this case) that should be used.

Specializing eq.~\pref{PCresult} to $E \simeq H$ then gives
\ba \label{PCresultH}
 \A_\E (E) &\simeq& H^2 M_p^2 \left( \frac{1}{M_p} \right)^\E
 \left( \frac{H}{4 \pi \, M_p}
 \right)^{2L} \prod_{d_n = 2} \Bigl( c_n \Bigr)^{V_n} \\
 && \qquad \qquad \qquad \times
 \prod_{d_n = 0} \Bigl( \lambda_n \Bigr)^{V_n}
 \prod_{d_n \ge 4} \left[ g_n \left( \frac{H}{M_p}
 \right)^2 \left( \frac{H}{M}
 \right)^{d_n-4} \right]^{V_n}  \,, \nn
\ea
where $\lambda_n \lsim \O(1)$ provided $V/v^4$ varies appreciably
only when $\phi$ changes by an amount of order $M_p$. This shows
the irrelevance of the $g_n$ terms (having 4 or more derivatives),
as well as the lack of additional suppression of the $d_n = 2$ and
$d_n = 0$ interactions, beyond the basic loop-suppression factor.

To apply this to the one-loop inflaton fluctuations themselves,
$\langle \phi^2 \rangle$, recall that the quartic interaction in
the scalar potential is $\lambda_4 v^4 (\phi/M_p)^4 \simeq
\lambda_4 (H/M_p)^2 \phi^4$. The one-loop graph involving this
vertex contributes an amount of order $\lambda_4 (H/M_p)^2 \langle
\phi^2 \rangle$ to the 2-point function, which can be compared
with eq.~\pref{PCresultH} specialized to $\E = 2$ to read off the
size of $\langle \phi^2 \rangle$. This gives the estimate
\be
 \langle \phi^2 \rangle \simeq \left( \frac{H}{4\pi} \right)^2 \,,
\ee
in agreement with the standard calculations. Indeed it is this
connection between $\langle \phi^2 \rangle$ and $H^2$ that is
responsible for the numerator of the observable combination
$\delta = H^2 / \dot \phi$ discussed earlier. (The $\dot\phi$
comes from the requirement that the $\phi$ fluctuation mix with
the metric to generate a curvature fluctuation that can be
observed in the CMB.)

\subsection{Applications}

As an example of the utility of these power-counting estimates we
apply the above reasoning to identify the domain of validity of
semiclassical methods in two closely related inflationary models.

\subsubsection{Higgs inflation}

Using this formalism, we now consider the example of Higgs inflation \cite{Hinfa} that has recently
gained some attention  \cite{Hinf,HinfRG}. This model starts with
the very economical proposal to try to obtain inflation using the
Standard Model Higgs as the inflaton. The idea is to do so by
supplementing the Standard Model and Einstein-Hilbert lagrangian
densities with the sole dimension-4 interaction that is not
normally written down\footnote{Earlier examinations of
non-minimally coupled models include \cite{earlierNMC}}:
\be
 \L_{H\,{\rm inf}} = \L_{\scriptscriptstyle SM} +
 \L_{\scriptscriptstyle EH} + \xi \, \H^\dagger \H
 \, R \,,
\ee
where $\H$ is the usual Standard Model doublet $\sqrt2 \, \H = (0,
v_\ssH + h)^{\scriptscriptstyle T}$, and $\xi$ is a dimensionless
coupling. In particular, the Higgs potential is the usual quartic
form,
\be
 V = \lambda_\ssH \left( \H^\dagger \H
 - \frac{v_\ssH^2}{2} \right)^2
 = \frac{\lambda_\ssH}{4} \left(2 \, v_\ssH h + h^2
 \right)^2 \,,
\ee
where $\lambda_\ssH$ is related to the Higgs boson mass by
$m_\ssH^2 \simeq 2 \lambda_\ssH  v_\ssH^2$. Because the rest of
the action is completely determined by non-inflationary physics,
the only adjustable parameter with which to try to make the model
inflate is $\xi$.

Once one performs a Weyl rescaling to transform to the Einstein
frame the Higgs potential becomes
\be
 V_\EF \simeq \frac{\lambda_\ssH (\H^\dagger \H - v_\ssH^2/2)^2}{
 (1 + \xi \, \H^\dagger \H / M_p^2)^2} \,,
\ee
which is to be regarded as being a function of $h(\phi)$, where
$\phi$ is the field that canonically normalizes the Einstein-frame
Higgs kinetic term. Remarkably, this can be flat enough to
inflate, provided that there is a reliable regime for which
$\H^\dagger \H \gg v_\ssH^2$ and $\xi \H^\dagger \H \gg M_p^2$,
since in this case $V_{\scriptscriptstyle EF} \simeq \lambda_\ssH
M_p^4/\xi^2$ is approximately constant.

More precisely, expressing the potential in terms of the canonical
variable in the inflationary regime gives
\be
 V_\EF \simeq \frac{\lambda_\ssH M_p^4}{\xi^2} \Bigl[
 1 + A e^{-a \phi/M_p} \Bigr]^{-2} \,,
\ee
where $A$ and $a$ are dimensionless numbers. The inflationary
regime of interest is then $\phi \gg M_p$, since in this case $[1
+ A e^{-a x}]^{-2} \simeq 1 -2 A e^{-ax} + \cdots$ is
approximately constant. Dropping $\O(1)$ constants, this shows
that the energy density during inflation is $V \simeq v^4$ where
$v^2 \simeq \sqrt{\lambda_\ssH} M_p^2/\xi$, and so the Hubble
scale during inflation is $H \simeq v^2/M_p \simeq
\sqrt{\lambda_\ssH} M_p/\xi$. Computing the value of the slow-roll
parameters at horizon exit and demanding $\delta \simeq 7  \times
10^{-4}$ then shows that the amplitude of primordial fluctuations
agrees with observations provided
\be
 \xi \simeq 5 \times 10^4 \, \sqrt{\lambda_\ssH}
 \simeq 5 \times 10^4 \left( \frac{m_{\scriptscriptstyle H}}{
 \sqrt 2 \; v_{\scriptscriptstyle H}} \right) \gg 1\,,
\ee
where $m_{\scriptscriptstyle H} > 115$ GeV and
$v_{\scriptscriptstyle H} = 246$ GeV respectively denote the mass
and expectation value of the Higgs \cite{Hinfa}.

The large size of the coupling $\xi$ is unusual from the particle
physics perspective, and leads one to worry about whether multiple
insertions of the corresponding effective interactions might
generate unexpectedly large quantum effects. This is the kind of
question for which the above power-counting arguments are well
suited. Although the size of some loop effects were examined in
refs.~\cite{Hinf,HinfRG}, the generality of the power-counting
result given earlier allows the effects of couplings to be
identified systematically.

Since the Einstein-frame potential energy varies by of order $v^4$
when $\phi$ ranges through the range $M_p$ the power-counting
result, eq.~\pref{PCresult}, may be directly applied. In
particular, it can be used to put an upper bound on the energy
scale, $M$, at which the low-energy effective description must
break down. This is most easily done by studying energetic
graviton-Higgs scattering, $g h \to g h$, or Higgs-Higgs
scattering, $h h \to h h$, in flat space, and asking when this
saturates the unitarity bound as a function of the loop order $L$.
For this purpose we may apply eq.~\pref{PCresult} to the
scattering amplitude, taking $E$ to be the center-of-mass energy
of the scattering. Furthermore, because we expand about flat space
and small Higgs vev we may regard the interaction $\xi \H^\dagger
\H R$ as an interaction vertex involving $d_n = 2$ derivatives.

To obtain the bound we concentrate on the potentially most
dangerous graphs that involve only the coupling $\xi$. According
to eq.~\pref{PCresult}, an $L$-loop graph of this type that
involves $V_n$ insertions of the $\xi$ coupling constant
contributes to the ($\E = 4$)-point amplitude an amount
\be \label{PCresultHinf}
 \A_4 (E) \simeq \left( \frac{E}{M_p} \right)^2
 \left( \frac{E}{4 \pi \, M_p}
 \right)^{2L}  \prod_n \xi^{V_n}
 \,,
\ee
where the product is over the power, $N_n$, of the fields $h$ and
$h_{\mu\nu} = g_{\mu\nu} - \eta_{\mu\nu}$ appearing in the
expansion of the original interaction $\xi \H^\dagger \H R$.

By virtue of the identity, eq.~\pref{checkID}, the quantities
$V_n$ and $N_n$ are related to $L$ and $\E = 4$ by $\sum_n (N_n -
2)V_n = \E - 2 + 2L = 2 + 2L$, and so the largest power of $\xi$
at any fixed loop order arises from multiple insertions of the
$N_n = 3$ vertex, in which case $V_{\rm max} = 2 + 2L$. The
highest power of $\xi$ appearing at any fixed order in $L$ then
becomes
\be \label{PCresultHinfL}
 \A_4^{\rm max} (E) \simeq \left( \frac{\xi E}{M_p} \right)^2
 \left( \frac{\xi E}{4 \pi \, M_p}
 \right)^{2L}
 \,.
\ee
At tree level this gives $ \A_{4,tree} \propto \xi^2$,
corresponding to the scattering graph involving two trilinear
$h-h-h_{\mu\nu}$ vertices. (Notice that for graviton-Higgs
scattering this is a stronger dependence than the linear
dependence in $\xi$ coming from the naive graph involving no
internal lines at all, that uses the quartic
$h-h-h_{\mu\nu}-h_{\lambda\rho}$ vertex, demonstrating the utility
of the power counting analysis.)

Demanding that the cross section built from a term like this not
saturate the unitarity bound, $\sigma \propto 1/E^2$, gives a
$\xi$-dependent upper bound on how large $E$ can sensibly be
within the low-energy theory, leading to
\be \label{EUB}
 E < E_{\rm max} \simeq \frac{M_p}{\xi} \,.
\ee
For Higgs-Higgs scattering through graviton exchange this
power-counting estimate reproduces the results of an explicit
calculation \cite{HW}, with the $\O(1)$ numerical factor not
written explicitly in eq.~\pref{EUB} revealed to be
$\sqrt{\pi/6}$. This provides a quantitative upper bound on the
true cut off of the theory.

Eq.~\pref{EUB} is useful because it furnishes an upper bound as to
how big the scale $M$ can be that controls the size of
higher-derivative terms in the low-energy effective
theory.\footnote{Notice that this upper bound for $M$ is
parametrically smaller than the value $M_p/\sqrt \xi$ sometimes
found in the literature. We believe this misidentification of the
unitarity bound in the literature is due to not basing it on the
strongest possible dependence on $\xi$.} Some new physics must
intervene at a scale $M < M_p/\xi$, so long as the more
microscopic underlying physics whose low-energy sector the
effective theory captures is itself unitary. Because the Hubble
scale is $H \simeq \sqrt{\lambda_\ssH} M_p/\xi$ in this picture,
the identification $M \lsim M_p/\xi$ implies $H/M \gsim
\sqrt{\lambda_\ssH}$. This leaves only the narrow window $1 \gg
H/M \gg \sqrt{\lambda_\ssH}$ within which all approximations
remain valid.\footnote{It should be remarked that $\lambda_\ssH
\simeq 0.03 \, \lambda_{\ssH 0}$ can be smaller than the value
$\lambda_{\ssH 0}$ relevant for Higgs physics at the LHC once it
is run up to the large energies relevant to inflation
\cite{HinfRG}, leading to $\sqrt\lambda_\ssH \simeq 0.2
\sqrt{\lambda_{\ssH 0}}$.}

This window gets more uncomfortable the more the new physics
couples to the Higgs field, since we've seen that the
approximately constant inflationary potential relies on there
being a regime for which $V \propto (\H^\dagger \H)^2$ and the
non-minimal coupling to gravity is $f R$ with $f \propto
\H^\dagger \H$. Although this is the case for quartic $V$ and
quadratic $f$ when $\H^\dagger \H \gg M_p^2/\xi \gg v_\ssH^2$, it
need no longer remain so once terms of order $\delta V \propto
(\H^\dagger \H)^3$ or $\delta f \propto (\H^\dagger \H)^2$ (or
higher) are generated by loops. Furthermore, as is seen from
eqs.~\pref{extree} and \pref{exloop}, these corrections
generically need not be small.

For instance, a quartic coupling of the form $g \H^\dagger \H \,
\chi^\dagger \chi$ between the Higgs and a heavy field $\chi$
having mass $M_\chi$ cannot be forbidden by any internal
symmetries and would generate loop contributions $\delta V \simeq
g^3 (\H^\dagger \H)^3/(4\pi M_\chi)^2$ and $\delta f \simeq g^2
(\H^\dagger \H)^2/(4 \pi M_\chi)^2$. The quartic term in $V$ can
only dominate if $g^3 \H^\dagger \H/(4\pi M_\chi)^2 \ll
\lambda_\ssH$ and similarly the quadratic term in $f$ dominates if
$g^2 \H^\dagger \H/(4 \pi M_\chi^2) \ll \xi$. Using $\H^\dagger \H
\gg M_p^2/\xi$ in these conditions shows that the scale,
$\Lambda$, suppressing higher powers of the Higgs field must
satisfy $\Lambda \simeq 4\pi M_\chi/g \gg M_p
\sqrt{g/(\lambda_\ssH \xi)}$ (for $V$) and $\Lambda \gg M_p/\xi$
(for $f$). For $g \simeq \lambda_\ssH$ the first of these shows
--- not surprisingly --- that $\Lambda$ must be
greater than the typical size of the Higgs field during inflation,
$\Lambda \gg M_p/\sqrt\xi$. The second shows that this bound does
not get worse than $\Lambda \gg M_p/\xi$, even if $g/\lambda_\ssH$
should be smaller than $1/\xi$. That is,
\be \label{LambdaLB}
 \Lambda \gg M_p \sqrt{\frac{g}{\lambda_\ssH \xi}} \quad
 \hbox{(if $g > \lambda_\ssH/\xi$)} \qquad \hbox{or} \qquad
 \Lambda \gg \frac{M_p}{\xi} \quad
 \hbox{(if $g < \lambda_\ssH/\xi$)}.
\ee

On the other hand, within this model it is the same mass scale,
$M_\chi$, that ultimately suppresses generic higher-derivative
terms in the effective action, since higher-curvature terms are
also generated at one loop of the form $R^3/(4 \pi M_\chi)^2$. We
see that the quantity $M$ in the effective theory obtained by
integrating out $\chi$ is of order $M \simeq 4 \pi M_\chi$, and so
unitarity requires $4 \pi M_\chi \ll M_p/\xi$, or
\be \label{LambdaUB}
 \Lambda \ll \frac{M_p}{g \xi} \,.
\ee
Consider now the separate cases $g < \lambda_\ssH/\xi$ and $g
> \lambda_\ssH/\xi$. If $g > \lambda_\ssH/\xi$ then conditions
\pref{LambdaLB} and \pref{LambdaUB} together require $1/(g\xi) \gg
\sqrt{g/(\lambda_\ssH \xi)}$, or $g^3 \ll \lambda_\ssH/\xi < g$.
On the other hand, if $g < \lambda_\ssH/\xi$ then \pref{LambdaLB}
and \pref{LambdaUB} together simply require $g \ll 1$. We see
explicitly in this example how any other particles must be kept
very heavy and/or strongly sequestered from the Higgs in order for
the inflationary mechanism to be viable.\footnote{The potential
danger of these interactions, and the potential necessity for
there to be a desert involving no such virtual particles up to
these large scales was already recognized in the original
literature.}
\subsubsection{Inflation from Curvature-squared Terms}

As our second application we next examine inflationary proposals
that are based on higher-curvature interactions \cite{RsqInf},
which represent a variation on the above theme. Consider to this
end the curvature-squared action
\be \label{RsqL}
 \L =  \sqrt{-g} \; \left[- \frac{M_p^2}{2} \, R + \zeta \, R^2
 \right] \,.
\ee

The Hubble scale can be most easily identified by exploiting the 
relationship between this theory and the Higgs-Inflation theory. This can be made clear
by rewriting the $R^2$ Lagrangian as a scalar-tensor model by performing a Hubbard-Stratonovich transformation and `integrating in' a scalar field of dimension one $\Phi$, as in
\be \label{RsqPhi}
 \L = \sqrt{-g} \; \left[ - \frac{M_p^2}{2} \, R - 2\,\alpha \,
 \Phi^2 \, R - \Phi^4 \right].
\ee
Performing the gaussian integral over $\Phi$ returns the
lagrangian density of eq.~\pref{RsqL}, with $\zeta = \alpha^2$.

The relation between this model and the one previously considered can be seen 
by performing a conformal transformation on this theory to the Einstein 
frame $g^E_{\mu\nu}=f(\Phi)g_{\mu\nu}$
with $f(\Phi)=1+4\alpha \, \Phi^2/\, M^2_{p}$ such that the Lagrangian becomes,
\be
{\cal L}= \sqrt{-g_E}\Big(-\frac{1}{2}M^2_{p} \,  R_E - \frac{3}{4}M^2_{p} \, \frac{f'(\Phi)^2}{f(\Phi)^2} \,(\partial_E \Phi)^2-V_E(\Phi)\Big)
\ee
where the Einstein-frame scalar potential is
\be
V_E(\Phi)=\frac{\Phi^4}{\Big(1+\frac{4\alpha}{M^2_p}\Phi^2\Big)^2}.
\ee

Further transforming to a canonical scalar field $\sigma$ through the field transformation
\be
\sigma = \sqrt{\frac{3}{2}} M_P  \ln \Big(1+4\alpha \, \Phi^2/\, M^2_{p}\Big), 
\ee
the Einstein-frame scalar potential becomes 
\ba
V_E(\sigma)= \frac{M^4_P}{16 \alpha^2} \Big[1 - {\rm exp}\Big( - \sqrt{\frac{2}{3}} \sigma/ M_P\Big) \Big]^2.
\ea

The inflationary analysis therefore proceeds much as in Higgs inflation before, with
inflation occurring for fields  $\Phi \gg M_{p}/2\sqrt{|\alpha|}$ or $\sigma\gg \sqrt{\frac{3}{2}} M_P$, where the
Einstein-frame scalar potential is of order $V_\EF \simeq 
\lambda M_p^4/(16\xi^2)= M_p^4/16\zeta$ and the Hubble scale is $H \simeq
\sqrt\lambda \, M_p/\xi \simeq M_p/4\sqrt\zeta$.
 Again, successful generation of primordial density fluctuations requires the
combination $\xi/\sqrt\lambda = 4 \, |\alpha| = 4 \, \sqrt\zeta$ to
be large, of order $10^4$.

To see when $\zeta \simeq 10^8$ begins interfering with the
semiclassical approximation we again use the power-counting
arguments of previous sections. There are two equivalent ways 
to determine the bounds on $E$ for this theory, one can
directly analyze the given lagrangian 
and calculate the cut off scale for graviton-graviton scattering, $g g \to gg$, using the $d_n = 4$ interactions
of the lagrangian density.
Repeating the arguments used for Higgs Inflation above leads to a problem with unitarity once the
scattering energies reach $E \simeq E_{\rm max} =
M_p/\zeta^{1/3}$. Alternatively, one can use the theory after the Hubbard-Stratonovich transformation
in the einstein frame and power-count. Note that one wishes to power count interactions with 
no external $\Phi$ fields as in this case $\Phi$ is an auxiliary field and not a real field as in Higgsflation. One can construct effective $d_n = 4$ interaction operators and then power count 
directly as before, again obtaining a cut off scale  $E \simeq E_{\rm max} =
M_p/\zeta^{1/3}$.

In either approach, we require that the scale $M$ controlling all
other powers of curvature not written explicitly in
eq.~\pref{RsqL} to satisfy $M \ll M_p/\zeta^{1/3}$. But using the
above expression for the inflationary Hubble scale, $H \simeq
M_p/\sqrt\zeta$ then shows that the ratio $H/M$ must satisfy $H/M
\gg \zeta^{-1/6} \simeq 1/20$. Again inflation requires $H/M$ to
be close to a breakdown of the adiabatic approximation that
underlies the understanding of eq.~\pref{RsqL} as part of a
low-energy effective theory, making any inflationary conclusions
drawn using it somewhat suspect.

\section{Conclusions}

Quantum corrections to the semi-classical approximation generally
employed of the inflation literature can be critical in
determining the viability of particular inflationary scenarios.
Indeed, it is the fact that quantum effects are {\em not}
completely negligible that underlies the possibility of explaining
primordial fluctuations in terms of quantum fluctuations of the
inflaton.

Although calculating quantum effects in non-renormalizable
theories like gravity may be unfamiliar, there is a well-defined
framework within which it may be done. This framework was
developed and tested against experiment using non-renormalizable
theories elsewhere in physics, and relies on the observation that
the semiclassical limit in such cases is controlled by a
low-energy approximation.

In this paper we have applied standard power-counting arguments
for such theories that allow one to easily quantify the domain of
validity of the classical approximation within any particular
model. Indeed, it is because many slow-roll models lie well within
the classical limit that justifies the belief that inflation can
reliably be predicted using the standard classical analyses.

However the same may not be true for more exotic inflationary models,
or for models of dark energy for that matter, almost all of which
are founded on a purely classical analysis. We believe it behooves
the proponent of any such a scenario to justify that validity of
the classical approximation, which should be viewed as one of the
hurdles any serious proposal must clear.

As an application of these techniques, we have examined the domain
of validity of the Higgs-Inflaton scenario, and find that its
semiclassical analysis is consistent only if the scale, $M$,
governing the low-energy approximation lies in the narrow range,
$M_p/\xi \gg M \gg \sqrt{\lambda_\ssH} M_p/\xi$, where $\xi$ is
the coefficient of the non-minimal Higgs-graviton interaction,
$\xi \H^\dagger \H \, R$, and $\lambda_\ssH = m_\ssH^2/(2 \,
v_\ssH^2)$ is the usual Standard Model Higgs quartic
self-coupling. Although it is a logical possibility that such a
scale exists, we argue that it is extremely unstable to the
existence of any small couplings between the Higgs and other heavy
particles. The situation is similar for curvature-squared
inflationary models, which also must push the adiabatic
approximation that is essential to regarding such
theories as well-behaved low-energy effective descriptions of any
sensible underlying microscopic dynamics.

\section*{Acknowledgements}
We thank Guillermo Ballesteros for comments on the manuscript.
This work was partially supported by funds from the Natural
Sciences and Engineering Research Council (NSERC) of Canada.
Research at the Perimeter Institute is supported in part by the
Government of Canada through NSERC and by the Province of Ontario
through MRI.

\end{document}